\documentclass[9pt,conference]{IEEEtran}

\usepackage[utf8]{inputenc}
\usepackage{graphicx} 
\usepackage{dblfloatfix}
\usepackage{float} 
\usepackage{subfig}
\usepackage{todonotes} 
\usepackage{wrapfig} 
\usepackage{rotating} 
\usepackage[export]{adjustbox} 
\usepackage{verbatim} 
\usepackage{mathtools} 
\usepackage{latexsym} 
\usepackage{listings} 
\usepackage{color}
\usepackage[linesnumbered]{algorithm2e} 
\usepackage{amsfonts} 
\usepackage{pifont}
\usepackage{textcomp} 
\usepackage[square,sort,comma,numbers]{natbib} 
\usepackage{amsmath} 
\usepackage{booktabs}
\usepackage{filecontents}

\usepackage[nolist]{acronym}
\newacro{AMPQ}{Advanced Message Queuing Protocol}
\newacro{BLE}{Bluetooth Low Energy}
\newacro{CSV}{Comma-separated values}
\newacro{DSL}{Domain-specific Language}
\newacro{EGL}{Epsilon Generation Language}
\newacro{EMF}{Eclipse Modeling Framework}
\newacro{EMOF}{Essential Meta Object Facility}
\newacro{GATT}{Generic Attribute Profile}
\newacro{GPML}{General-purpose Modeling Language}
\newacro{DSML}{Domain-specific Modeling Language}
\newacro{IoT}{Internet of Things}
\newacro{MDSD}{Model-driven Software Development}
\newacro{MOCAP}{Micro-Ontology Context-Aware Protocol}
\newacro{MQTT}{MQ Telemetry Transport}
\newacro{OWL}{Web Ontology Language}
\newacro{QUDT}{Quantities, Units, Dimensions and Data Types}
\newacro{RDF}{Resource Description Framework}
\newacro{SSN}{Semantic Sensor Network}
\newacro{SWoT}{Semantic Web of Things}
\newacro{SWT}{Semantic Web Technology}
\newacroplural{SWT}[SWTs]{Semantic Web Technologies}
\newacro{SOSA}{Sensor, Observation, Sample, and Actuator}
\newacro{SPARQL}{SPARQL Protocol and Query Language}
\newacro{UML}{Unified Modeling Language}
\newacro{WSN}{Wireless Sensor Network}
\newacroplural{WSN}[WSNs]{Wireless Sensor Networks}
\newacro{WoT}{Web of Things}

\definecolor{paletteDarkBlue}{rgb}{0.5078125,0.76171875,0.92578125}
\definecolor{paletteMidBlue}{rgb}{0.69921875,0.88671875,0.94140625}

\lstdefinelanguage{Turtle}
    {morekeywords={DEG_C,float,Temperature,HumidityTempSensor,hasFeatureOfInterest,hasResult,madeBySensor,observedProperty,numericValue},
    otherkeywords={:unit},
    sensitive=false,
    morestring=[s]{"}{"},
    stringstyle=\color{black}\bfseries
}

\lstdefinestyle{mystyle}{
    morestring={*[d]{'}},
    commentstyle=\color{gray},
    stringstyle=\color{black},
    morekeywords={STRING,ID,INT},
    keywordstyle=\color{black}\bfseries,
    basicstyle=\ttfamily\scriptsize,
    breakatwhitespace=false,         
    breaklines=false,                 
    captionpos=b,                    
    keepspaces=true,                 
    showspaces=false,                
    showstringspaces=false,
    showtabs=false,                  
    tabsize=1,
    breaklines=true,
    postbreak=\mbox{{$\hookrightarrow$}\space}
}

\usepackage{textcomp}

\usepackage{tikz}
\newcommand*\circled[1]{\tikz[baseline=(char.base)]{
            \node[shape=circle,draw,inner sep=1pt] (char) {#1};}}
            
\usepackage{enumitem} 
\usepackage{balance}

\usepackage{amssymb}
\usepackage{pifont}
\usepackage{eso-pic} 
\usepackage[hidelinks]{hyperref} 
\newcommand{\cmark}{\ding{51}}%
\newcommand{\xmark}{\ding{55}}%

\newcommand{\frameworkname}{\textsc{Lemons}}

\newcommand{\IEEEAcceptedManuscriptNotice}{%
  \AddToShipoutPictureFG*{%
    \AtPageLowerLeft{%
      \put(48,9){%
        \parbox[b]{516pt}{%
          \hrule
          \vspace{2pt}
          \fontsize{5.2}{5.8}\selectfont
          \raggedright
          \textcopyright~2020 IEEE. Personal use of this material is permitted.
          Permission from IEEE must be obtained for all other uses, in any current
          or future media, including reprinting/republishing this material for
          advertising or promotional purposes, creating new collective works, for
          resale or redistribution to servers or lists, or reuse of any copyrighted
          component of this work in other works.

          \textit{This is the accepted manuscript of: J. Novacek, A. Kühlwein,
          S. Reiter, A. Viehl, O. Bringmann, and W. Rosenstiel, ``LEMONS:
          Leveraging Model-Based Techniques to Enable Non-Intrusive Semantic
          Enrichment in Wireless Sensor Networks,'' in 2020 46th Euromicro
          Conference on Software Engineering and Advanced Applications (SEAA),
          pp. 561--568, 2020. The version of record is available at
          \href{https://doi.org/10.1109/SEAA51224.2020.00092}{%
          \textcolor{blue}{https://doi.org/10.1109/SEAA51224.2020.00092}}.}%
        }%
      }%
    }%
  }%
}

\begin{document}
\IEEEAcceptedManuscriptNotice

\title{\frameworkname{}: Leveraging Model-Based Techniques \\to Enable Non-Intrusive Semantic Enrichment\\ in Wireless Sensor Networks}

\author{\IEEEauthorblockN{Jan Novacek\IEEEauthorrefmark{1}\IEEEauthorrefmark{2}, Arthur Kühlwein, Sebastian Reiter\IEEEauthorrefmark{1},\\
Alexander Viehl\IEEEauthorrefmark{1}, Oliver Bringmann\IEEEauthorrefmark{1}\IEEEauthorrefmark{2}, Wolfgang Rosenstiel\IEEEauthorrefmark{1}\IEEEauthorrefmark{2}}
\IEEEauthorblockA{\IEEEauthorrefmark{1}FZI Research Center for Information Technology\\
Haid-und-Neu-Str. 10-14, 76131 Karlsruhe}
\IEEEauthorblockA{\IEEEauthorrefmark{2}University of Tübingen\\
Sand 14, 72076 Tübingen}}

\maketitle              

\begin{abstract}
The paper presents an efficient approach to the semantic enrichment of measured sensor data in \acp{WSN}, by bridging techniques from \ac{MDSD} and \ac{SWT}. 
Our approach reinforces data interoperability, fostering data sharing and reuse, by utilizing \ac{SWT}.
Model-based and type-agnostic configuration reduces the overall effort for \ac{WSN} setup and maintenance, which are traditionally complex and time-consuming tasks.
The presented approach addresses the problem of large-scale \ac{WSN} management through the application of \ac{SWT} in \ac{WSN} configuration and management without requiring expert knowledge.
Additionally, we present a generic architecture and an implementation which is also supplemented by hands-on descriptions of an illustrative use case.
Our experimental results demonstrate that our model-based approach provides non-intrusive semantic enrichment with sub-millisecond computational overhead, as well as partially automated configuration of \acp{WSN}.\\
\end{abstract}

\section{Introduction}
\label{sec:introduction}
The \ac{IoT} paradigm promises the interconnection of virtual and physical objects, with applications bringing considerable benefits to industry, healthcare, and other domains.
It is estimated that currently 1.5 trillion "things" exist, of which about 99$\,$\% are projected to likely become part of a network \cite{greengard2015internetOfThings}.
\acfp{WSN} as a key \ac{IoT} technology are increasingly applied in a wider range, with estimations putting the global \ac{WSN} market at \$1.8 billion by 2024 \cite{idtechex}.
To establish a \ac{WSN}, interoperability needs to be facilitated between the network devices which typically exhibit heterogeneity in formats, domains, and types.

This heterogeneity makes data provisioning and data interpretability a major challenge \cite{shi2018iotSurvey}.
Furthermore, the management of \ac{WSN} applications has been reported to be a key issue \cite{yick2008wsnSurvey, mostafaei2018softwareWSNSurvey}.
For large-scale \acp{WSN} with thousands of heterogeneous nodes, network configuration becomes challenging and can be a time-consuming, complex, and error-prone task \cite{netconf_thesis}.
The application of semantic technologies to the \ac{IoT} addresses the former issue by enabling device interoperability and facilitating interpretability of sensor data \cite{shi2018iotSurvey}.
In particular, ontological representations of \acp{WSN} can aid in the provisioning of measurement data, for instance by providing descriptions with explicit semantics.
However, the creation of corresponding ontology instances is also affected by the aforementioned scalability issues and requires the knowledge of a domain expert.
\begin{figure}
  \centering
  \includegraphics[width=\linewidth]{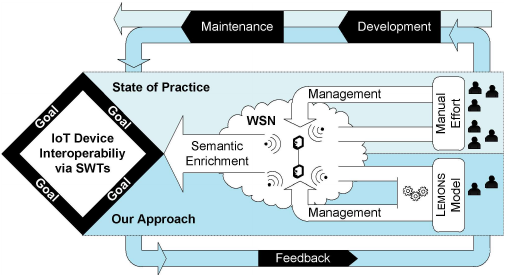}
  \caption{Comparison of the state of practice and our approach}
  \label{fig:approach-comparison}
\end{figure}
In addition to that, \ac{WSN} configuration is highly dependent on application objectives and performance requirements \cite{ibrahim2014wsnConfigurationInConstruction}. Consequently, situation-specific considerations are required for \ac{WSN} configuration.
Furthermore, there exist \ac{WSN} applications which require continuous adaption, for instance in construction \cite{ibrahim2014wsnConfigurationInConstruction}.

To address the mentioned issues we introduce \frameworkname{}\footnote{\textbf{Le}veraging \textbf{Mo}del-Based Techniques to Enable \textbf{N}on-Intrusive \textbf{S}emantic Enrichment in Wireless Sensor Networks}, an approach leveraging techniques from \acf{MDSD} and \acf{SWT} to reduce complexity in development and maintenance processes and improve overall quality. In particular, this paper presents a model-based approach to improve the efficiency of \ac{WSN} management and the enhancement of sensor data with explicit semantics.
An inter-wining combination of model-based code generation for \ac{WSN} gateways and the consequent use of standardized communication protocols, data formats and \ac{OWL} ontologies reduces complexity in development and maintenance processes and improves overall quality by generating 100$\,$\% of \frameworkname' application code.
The approach bridges the \ac{MDSD} and \ac{SWT} techniques by providing abstractions of ontological constructs in the metamodel.
It is applicable to any kind of \ac{WSN} independent of the protocols and devices used and is highly suitable for the integration into DevOps-oriented \cite{mala2019integratingIoTandSE} processes as \ac{WSN} management capabilities enabled by our approach can be used in development, setup and maintenance phases as well. Moreover, feedback from operation can directly be considered in further development.
Fig. \ref{fig:approach-comparison} outlines the envisioned approach.

The main contribution of this paper is the application of \ac{SWT} in large-scale \ac{WSN} configuration and management without requiring expert knowledge to address the problem of large-scale \ac{WSN} application management.

\begin{figure}
  \centering
  \includegraphics[width=\linewidth]{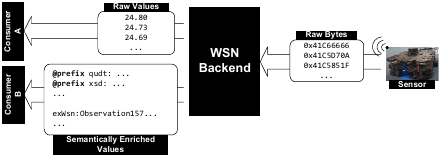}
  \caption{Overview of the semantic enrichment process for our use case}
  \label{fig:approach-overview-use-case}
\end{figure}

The primary benefits of our approach are:
\begin{itemize}[align=left,labelwidth=2cm,labelindent=15pt,leftmargin=2.15cm]
    \item[Efficiency]{
        \begin{itemize}
            \item Reduction of \ac{WSN} setup effort and complexity
            \item Acceleration of the \ac{WSN} deployment 
            \item Easy extension and adjustment of \acp{WSN}
            \item Development and maintenance integration
        \end{itemize}
    }
    \item[Quality]{
        \begin{itemize}
            \item Reduced setup and deployment error-proneness
            \item Better \ac{WSN} maintainability and reusability
        \end{itemize}
    }
    \item[Interoperability]{
        \begin{itemize}
            \item Improvement of measurement data interpretability
            \item Applicability to a board range of \acp{WSN}
        \end{itemize}
    }
\end{itemize}
We demonstrate and evaluate our proposed approach in a use case, where we enrich sensor network measurements with semantic information, see Fig. \ref{fig:approach-overview-use-case}.
In this use case, a mobile robot platform was equipped with a wireless sensor node measuring various features via different sensors.
The \ac{WSN} backend was processing received raw bytes from the sensor node and provisioned measurement data as raw values as well as a semantically enriched variant.

The remainder of this paper is structured as follows:
Section \ref{sec:related_work} provides relevant information on the background and related work.
Section \ref{sec:approach} describes the approach while Sections \ref{sec:use_cases} and \ref{sec:evaluation}, the use case and evaluation results are presented, respectively.
Finally, a conclusion drawn from the experiences made while carrying out this research is given in Section \ref{sec:conclusion}.

\section{Background and Related Work}
\label{sec:related_work}
This section provides brief introductions to fundamental technologies namely \acp{WSN}, the \ac{SWoT} and \ac{MDSD}. In addition to that, we compare our approach to related work.

\subsection{\aclp{WSN}}
A \ac{WSN} is an ad-hoc network consisting of sensor nodes and at least one so-called \emph{sink} acting as a data consumer or a bridge to another network \cite{akyildiz2002wireless,djedouboum2018bigWSN}. In contrast to other networks, \acp{WSN} are designed for specific applications which may be a reason that application management is an issue in \acp{WSN} \cite{yick2008wsnSurvey}.
The range of \ac{WSN} applications is broad, including environmental, health, smart home and industrial respectively commercial applications \cite{akyildiz2002wireless}.


A related approach is presented by Ibrahim and Moselhi in \cite{ibrahim2014wsnConfigurationInConstruction}. This approach has a focus on the design and configuration of \ac{WSN} hardware and software in the context of construction. The approach proposes an iterative methodology which includes rapid-prototyping.
The resulting system is self-adaptive with regard to contextual relevance monitoring.
While this approach also focuses on \ac{WSN} configuration, the solution concept is fundamentally different to our approach as it does neither consider \ac{MDSD} nor \ac{SWT} techniques and is also not concerned with the data provisioning aspect.

\subsection{\acl{SWoT}}
The \ac{SWoT} is focused on providing interoperability among the heterogeneous "things" in the \ac{IoT} \cite{jara2014semantic}.
Typical applications include Smart Homes, E-Health and Smart Cities \cite{shi2018iotSurvey}.
There is a large body of research concerned with the creation of ontologies for semantic interoperability \cite{schlenoff2013literature,gyrard2018survey,nagowah2018overview}.
With regard to ontologies in \acp{WSN} the \ac{SSN} ontology plays a key role \cite{ganzha2016semanticIoT}.
This ontology is used to model sensors, the data they produce as well as actuators.
Moreover, it allows the description of used procedures for measurements and generating samples and the description of features of interest.
It has a modular design and consists of the core ontology named \ac{SOSA} and the upper \ac{SSN} ontology.
Because of its popularity, availability and standardization we also decided to reuse the \ac{SSN} ontology.

Bermudez et al. recently presented in \cite{bermudez2016iotLite} \emph{IoT-Lite}, which is also an \ac{SSN}-based ontology.
The authors follow the principle that \ac{IoT} ontologies need to be lightweight due to the typical resource limitations of \ac{IoT} devices and the demand for low processing latencies.
Also, the authors refer to a work by Kolozali et al. in \cite{kolozali2014IoTAnnoations} which presents a framework for real-time semantic annotation of \ac{IoT} data. While being similar to our approach, this framework builds upon \ac{AMPQ} and is therefore dependent on a certain protocol.
Another related work is the proposal of the \ac{MOCAP} by Sahlmann and Schwotzer in \cite{sahlmann2015mocap} which introduced the idea of using micro-ontologies in the communication between sensors and other devices. While we also use micro-ontologies in our approach, we instead employ them on the layers above the individual sensors in the network hierarchy.

\subsection{\acl{MDSD}}
\acf{MDSD} is a technique for developing complex software by using different models which provide suitable abstractions over particular aspects of the software that is being developed.
Besides analysis and optimization, these models can also be used to partially or fully generate the software code.

\ac{MDSD} has been used successfully in a number of different domains \cite{mdesurvey}, and the body of existing graphical and textual modeling languages is comprehensive, ranging from general-purpose modeling languages to \acfp{DSML}.
In the case of textual \acp{DSML}, we refer to them as \acfp{DSL}.
While graphical and textual representations form the concrete syntax of a modeling language, the metamodel to which it conforms is its abstract syntax.

The \acf{EMF} \cite{emfbook} is an Eclipse-based framework that enables the development of modeling languages conforming to Ecore, which is the meta-metamodel provided by \ac{EMF}.
That is, Ecore provides the basic constituents from which the metamodel of a modeling language can be constructed.
A large number of mature frameworks and robust tools have been developed around the \ac{EMF} that allow for the rapid development of graphical and textual modeling languages, including the corresponding editors.

In research, there has been some interest in the application of \ac{MDSD} to \acp{WSN}.
For instance, Beckmann and Thoss \cite{mde_dds_wsn} propose an \ac{MDSD} approach leveraging the Data Distribution Service for Real-time Systems (DDS) for the runtime and communication infrastructure to expedite the software development process.
Approaches such as \cite{wsn_mgmt_1}, \cite{wsn_mgmt_2}, and \cite{wsn_mgmt_3} propose the utilization of custom modeling languages to address the development, management, analysis, and simulation of \acp{WSN}.
However, none of these works directly address semantic aspects, such as the enrichment of measurement data with explicit semantics.

\begin{figure*}
  \centering
  \includegraphics[width=0.6\linewidth]{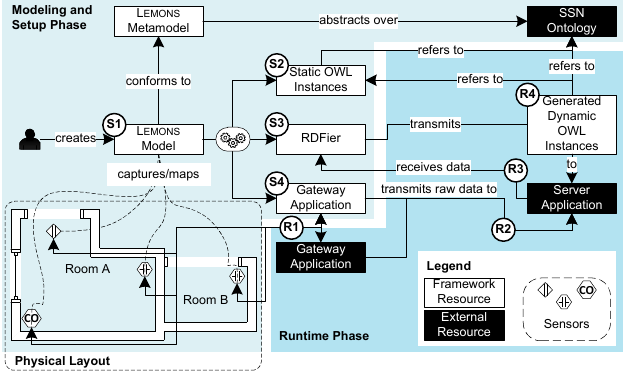}
  \caption{Overview of the overall approach}
  \label{fig:approach-overview}
\end{figure*}

\section{Approach}
\label{sec:approach}
This section describes our approach starting with an overview.
Section \ref{sec:metamodel} describes the \frameworkname{} metamodel.
Section \ref{sec:semantic-enrichment} elaborates on the the difference between \emph{static} and \emph{dynamic} OWL instances in semantic enrichment.
Section \ref{sec:system-architecture} then provides a description of our proposed system architecture while Section \ref{sec:implementation} completes the description of the approach with details on the implementation.

\subsection{Overview}
The approach is mainly structured in two phases: the \ac{WSN} \emph{modeling and setup phase} and the \emph{runtime phase}. Please refer to Fig. \ref{fig:approach-overview} for an overview.
During \emph{\ac{WSN} setup} a developer creates a \frameworkname{} model of the whole \ac{WSN} structure including representations of gateways and connected sensors along with the corresponding semantics \circled{\scriptsize{S1}}. This model conforms to a predefined metamodel which itself abstracts over the \ac{SSN} ontology. While the creation of this model is supported by a modeling suite, it is still a manual task during which knowledge from a number of sources has to be processed.
Based on the created model, three artifacts are generated: Static \ac{OWL} instances describing used sensors and platforms as well as observable properties and features of interest \circled{\scriptsize{S2}}, the \emph{RDFier} component which is responsible for the generation of dynamic \ac{OWL} instances at runtime later on \circled{\scriptsize{S3}} and optionally an executable for the \ac{WSN} gateway \circled{\scriptsize{S4}}. The two preceeding components are described in more detail in Section \ref{sec:implementation}.

At \emph{\ac{WSN} runtime}, the \ac{WSN} gateway will receive sensor data \circled{\scriptsize{R1}}.
Next, the gateway will transmit this measured sensor data to a server application \circled{\scriptsize{R2}}.
The server application is responsible for the distribution of the measured sensor data and will send it to the \emph{RDFier} component \circled{\scriptsize{R3}} which will in turn perform semantic enrichment through the generation of dynamic \ac{OWL} instances. At last, semantically enriched measurement data is transmitted back to the server \circled{\scriptsize{R4}} from which it can be distributed and processed further.

\subsection{Metamodel}
\label{sec:metamodel}
\begin{figure}
  \subfloat[Simplified domain diagram of the \frameworkname{ }metamodel]{\includegraphics[width=\linewidth]{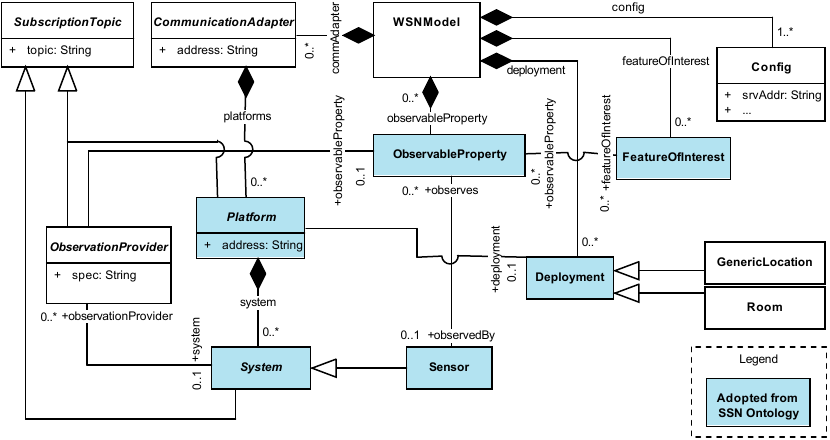}}\\
  \captionsetup[subfigure]{justification=justified,singlelinecheck=false}
  \subfloat[Grammar of the sensor data interpretation \ac{DSL} in W3C grammar notation]{
\begin{lstlisting}[frame=single,basicstyle=\tiny]
Model    ::= CharSpecExpr
CharSpecExpr
         ::= ID ':=' SpecExpr
SpecExpr ::= SimpleSpecExpr
           | FloatSpecExpr
           | CompoundSpecExpr
SimpleSpecExpr
         ::= TypeExpr ( 'unit' STRING ( 'factor' Number )? )? ( 'type' STRING )? ( 'property' ID )?
FloatSpecExpr
         ::= TypeExpr '.' TypeExpr ( 'unit' STRING ( 'factor' Number )? )? ( 'type' STRING )? ( 'property' ID )?
TypeExpr ::= UIntExpr
           | IntExpr
           | QuatExpr
CompoundSpecExpr
         ::= '(' CompoundSpecElementExpr ( ',' CompoundSpecElementExpr )* ')'
CompoundSpecElementExpr
         ::= SpecExpr 'as' 'topic'? ID ( 'unit' STRING ( 'factor' Number )? )? ( 'type' STRING )? ( 'property' ID )?
UIntExpr ::= 'UInt' '(' INT ')'
IntExpr  ::= 'Int' '(' INT ')'
QuatExpr ::= 'Quat' '(' INT ',' INT ')'
Number   ::= '-'? INT
\end{lstlisting}} 
  \caption{Constituents of the \frameworkname{} metamodel}
  \label{fig:approach-metamodel}
\end{figure}
\begin{figure*}[!t]
  \centering
  \includegraphics[width=0.9\linewidth]{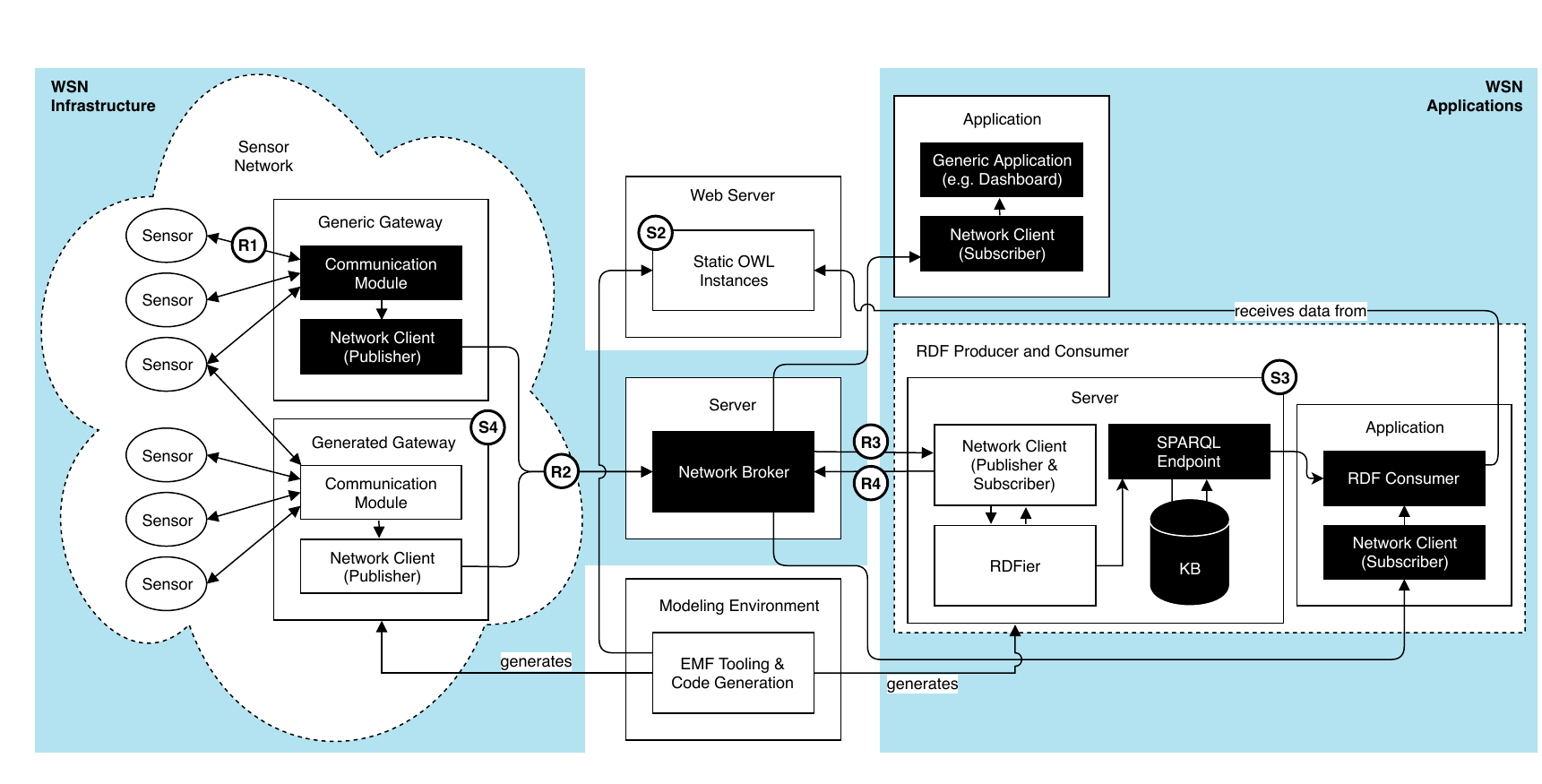}
  \caption{Overview of the system architecture}
  \label{fig:system-architecture}
\end{figure*}
Fig. \ref{fig:approach-metamodel} (a) depicts a simplified domain diagram of the \frameworkname{} metamodel, which provides the core concepts to model the physical and logical configuration of the \ac{WSN}.
In addition, the metamodel captures information required for expressing explicit semantics for the description of the \ac{WSN}, including measurement data provided by the sensors.

While one of the major design goals for the metamodel was technology agnosticism in terms of modeling the \ac{WSN} structure, we used state-of-the-art technologies as inspiration.
For instance, the \emph{SubscriptionTopic} element is related to the concept of topics in contemporary publish/subscribe messaging systems.
Integration of specific technologies in the \frameworkname{} metamodel is facilitated by specializing these core concepts.
An example of this integration is given in the use cases in Section \ref{sec:use_cases}.

The metamodel also abstracts over key concepts of the \ac{SSN} ontology with the aim of semantically enriching measurement data.
The development of these abstractions cannot be easily automated due to the complexity of the \ac{SSN} ontology and the differences between \ac{OWL} and Ecore in terms of expressiveness, as summarized in \cite{emf2rdf}.
Thus, we follow a manual, iterative development strategy consisting of two phases.
First, we identify and carefully analyze aspects of the \ac{SSN} ontology which are central for our approach.
These aspects are then subsumed into suitable abstractions, if possible, and integrated into the \frameworkname{} metamodel.
The parts of the \frameworkname{ }metamodel which have been adopted from the \ac{SSN} ontology are highlighted accordingly in Fig. \ref{fig:approach-metamodel} (a).

All information about the configuration of a \ac{WSN} is captured in the root-level element \emph{WSNModel}.
The \emph{Config} element holds configuration data for connecting with the network broker.

In addition to the elements depicted in Fig. \ref{fig:approach-metamodel} (a), the metamodel comprises a textual \ac{DSL} that allows for the specification of how individual bytes of raw sensor data are to be interpreted.
This specification is captured in the \emph{spec} attribute of the \emph{ObservationProvider} element.
Fig. \ref{fig:approach-metamodel} (b) depicts the grammar of the \ac{DSL} in W3C grammar notation\footnote{https://www.w3.org/TR/xquery-31/\#EBNFNotation}.
These specifications are used by code generators to construct the correct messages for data transmission.
Furthermore, the specifications assign semantics to each piece of raw measurement data in terms of the observable properties defined in a given \frameworkname{} model.

\subsection{Semantic Enrichment}
\label{sec:semantic-enrichment}
We used the \ac{SSN} ontology \cite{compton2012ssn} in our approach to describe sensors, sensor platforms and measured data.
With regard to measurement data units and types we used the \ac{QUDT} ontologies \cite{hodgson2014qudt}.
For expressing facts related to time we used the Time ontology \cite{hobbs2006time}.
All knowledge with which the raw sensor measurement data is enriched stems from the \frameworkname{} model.

In our approach we distinguish between \emph{static} and \emph{dynamic \ac{OWL} instances} - both are \emph{\ac{OWL} individuals} representing relevant \ac{WSN} information.
When we refer to \emph{static \ac{OWL} instances}, we mean the descriptions of the \ac{WSN} which do not change very frequently e.g. connected sensor nodes and their sensors or the overall network structure. Please note, that this does not mean, that corresponding \ac{OWL} instances representing such information cannot be changed over time.
In contrast to that, \emph{dynamic \ac{OWL} instances} carry volatile information such as sensor measurement data.
A small set of these dynamic instances forms a \emph{micro-ontology}.
Referring to \ac{SSN} terms, \emph{sensors}, \emph{platforms}, \emph{deployments}, \emph{systems} and \emph{features of interest} correspond to \emph{static} instances in our approach, while \emph{observations} correspond to \emph{dynamic} instances.
There are several reasons for this distinction: As static information does not change frequently it is beneficial to not transmit it over and over again thus saving bandwith in typically resource-constrained \acp{WSN}.
Besides that, when publishing static information in an ontology on a web server instead of constantly transmitting it increases the chances that it is available when needed.
This also makes a complicated query mechanism unnecessary.

While the static instances are generated directly from the \frameworkname{} model, dynamic instances are generated by the \emph{RDFier} component, see Fig. \ref{fig:approach-overview}.
Please see List. \ref{lst:sensor-platform-definition} to get an impression of an excerpt of the generated static instances and the bottom listing in Fig. \ref{fig:csv-vs-owl} for a generated dynamic instance.
The \emph{RDFier} component is in turn generated by the \frameworkname{} model.
In broad terms, the code generation creates the corresponding structures and expressions in the \emph{RDFier} source code which handle each of the various sensors described in the \frameworkname{} model.
The basic capabilities of the \emph{RDFier} comprise network connectivity to the server application, generation of \emph{observations} from received measurement data and publication thereof via the server or to a \ac{SPARQL} endpoint.

\subsection{System Architecture}
\label{sec:system-architecture}
Fig. \ref{fig:system-architecture} gives an overview of the system architecture.
The sensor network consists of multiple gateways, which are connected to sensor nodes.
The gateway is responsible for facilitating communication with the connected sensor nodes and the server.
The client application running on the gateway publishes the raw sensor data to the network broker running on the server, which is responsible for distributing the data to the applications that require it.

One of these applications is a server hosting the RDFier component, which consumes the raw sensor data and publishes semantically enriched versions of them to the network broker.
In addition, these enriched data are sent to a \ac{SPARQL} endpoint which is also running on the server.
Consequently, applications that require enriched data have the option of subscribing to a stream of data from the network broker or querying the \ac{SPARQL} endpoint.

The modeling environment contains the \ac{EMF} tooling for creating and editing \frameworkname{ }models as well as the corresponding generators for the gateway application and the server hosting the RDFier and \ac{SPARQL} endpoint.

Our decision to use this particular architecture is motivated by a number of considerations.
Because full access to the server cannot be assumed a priori, our approach required non-intrusiveness with regard to the existing \ac{WSN}.
Consequently, we decided to implement the RDFier component as another publisher and subscriber running independently of the rest of the system, effectively making the RDFier an extension.
In addition, placing the RDFier component inside the gateway applications would introduce communication overhead on the gateways running these applications, potentially decreasing overall network responsiveness.



\subsection{Implementation}
\label{sec:implementation}
\begin{figure}
  \centering
  \includegraphics[width=0.8\linewidth]{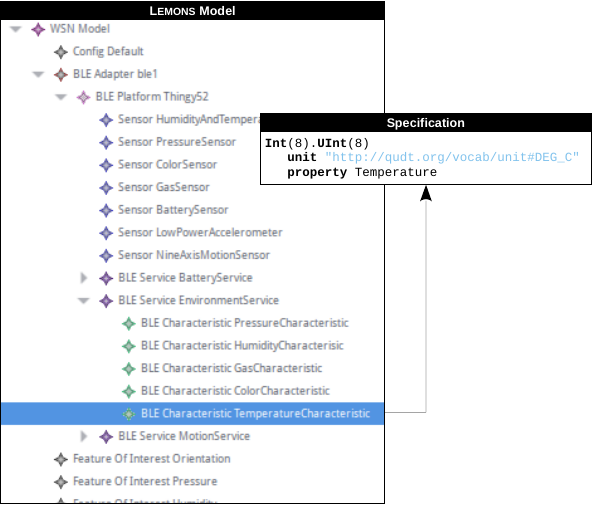}
  \caption{\ac{EMF} editor for our \frameworkname{} model with an exemplary specification of raw temperature measurement data}
  \label{fig:wsn-editor}
\end{figure}

As shown in Fig. \ref{fig:approach-overview}, a number of artifacts \circled{\scriptsize{S2}} to \circled{\scriptsize{S4}} are generated from a \frameworkname{ }model \circled{\scriptsize{S1}}, which can currently be viewed and edited using the default tree-based graphical editor that is generated by \ac{EMF} from the \frameworkname{ }Ecore description.
The graphical editor can be seen in Fig. \ref{fig:wsn-editor}.
While the generation of the static OWL instances \circled{\scriptsize{S2}} is handled by the Jena framework\footnote{https://jena.apache.org/}, \ac{EGL}\footnote{https://www.eclipse.org/epsilon/doc/egl/} templates are used to generate 100$\,$\% of the code for the RDFier \circled{\scriptsize{S3}} and gateway application \circled{\scriptsize{S4}}.

\setlength{\belowcaptionskip}{-10pt}

The gateway application is implemented within a single Scala \cite{odersky2004overview} object.
A basic design principle was the use of non-blocking I/O which we achieved through asynchronous stream-based processing utilizing the streaming engine Akka Streams\footnote{https://doc.akka.io/docs/akka/current/stream/index.html}.
Using Akka Streams had the advantage that we could easily define different \emph{sources} representing the sensors sending measurement data to the gateway depending on the type of the communication module.
Likewise, we could define various \emph{sinks} which represent the connections to the \emph{network broker} back-end consuming the raw measurement data, see \circled{\scriptsize{R2}} in Fig. \ref{fig:approach-overview} and Fig. \ref{fig:system-architecture}.
This pattern provides a great amount of maintainability because source and sink implementations could be reused and extended.
Also, it favors the template-based code generation.

For the RDFier application, which is also implemented in Scala, we applied the same pattern as in the gateway application implementation using the Akka Streams engine.
The main purpose of the RDFier component is the generation of \ac{SSN} ontology \emph{observations} as \ac{RDF} data based on received measurement data.
Therefore, the RDFier implementation has to contain code that transforms received raw measurement data into corresponding instances expressed in \ac{OWL}.
In a configuration for a large WSN there will be a lot of code constructs for each sensor value.
This is also where the code generation comes in: Based on the specified sensors of the \ac{WSN} in the \frameworkname{} model, constructs handling each sensor value transformation are automatically generated, thus potentially saving a considerable amount of time and avoiding errors.


\section{Use case}
\label{sec:use_cases}
This section presents an illustrative use case which has a focus on the provisioning of semantically enriched measurement data. While the presented approach is essentially applicable to any type of \ac{WSN}, in our use case the realizing system is relying on \ac{MQTT} \cite{banks2014mqtt} as communication protocol between the Raspberry Pi 3 Model B+ gateway, generic applications and the \emph{RDFier} and \ac{BLE} between the gateway and the sensor nodes.

\begin{figure}
  \centering
  \includegraphics[width=0.8\linewidth,frame]{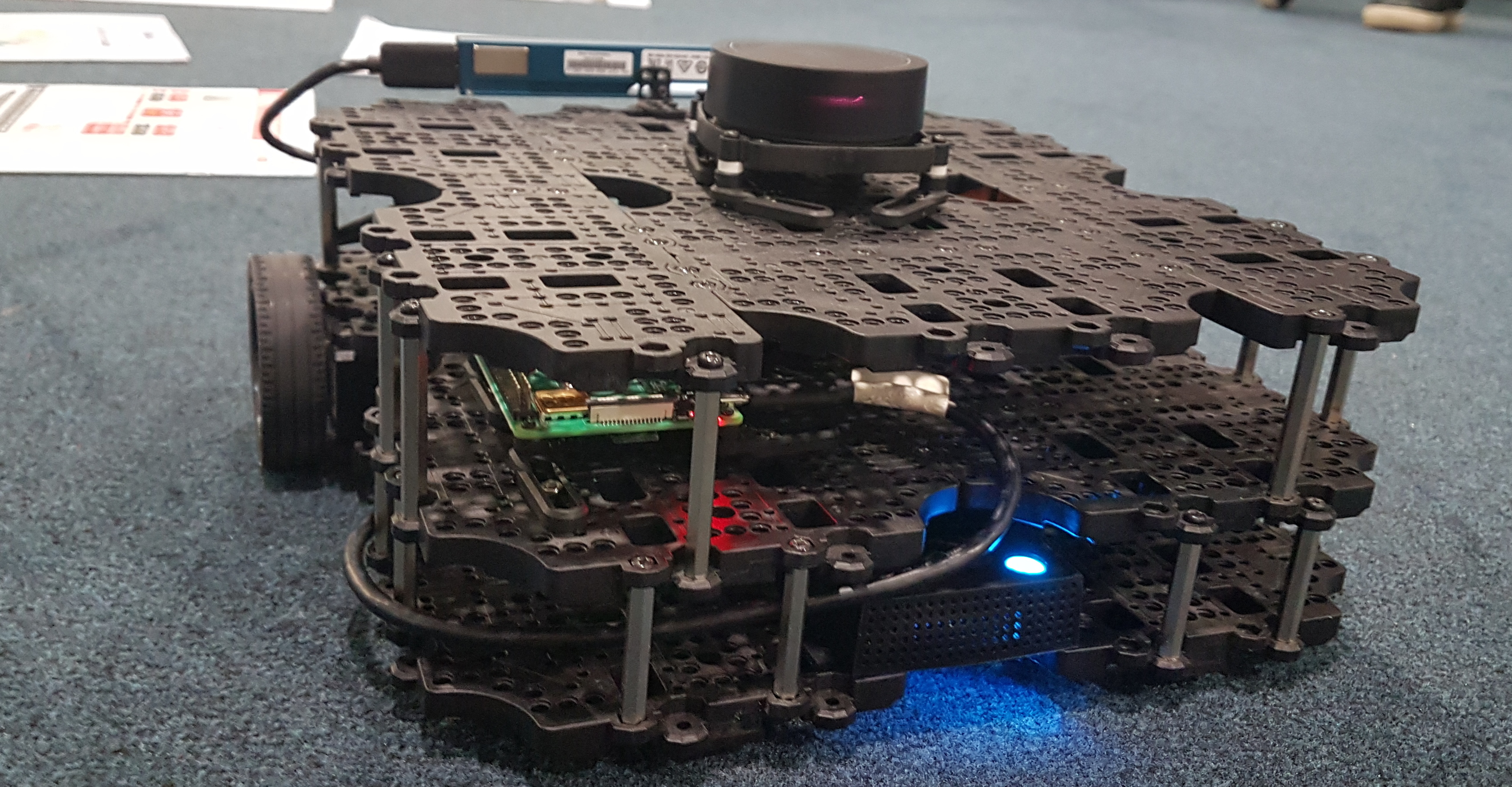}
  \caption{TurtleBot equipped with the Nordic Thingy:52\texttrademark{} \ac{IoT} sensor platform. The sensor platform is mounted on the lower middle in the back side of the robot}
  \label{fig:turtlebot}
\end{figure}

In order to generate load- and usage-profiles we equipped the Waffle TurtleBot\footnote{https://www.turtlebot.com/} with an additional \ac{IoT} sensor platform, the Nordic Thingy:52\texttrademark{}, see Fig. \ref{fig:turtlebot}.
The general setup of the use case is illustrated in Fig. \ref{fig:approach-overview-use-case}, where the backend comprises the network broker and the server \circled{\scriptsize{S3}} in Fig. \ref{fig:system-architecture}.

\begin{figure}
  \centering
  \includegraphics[width=\linewidth]{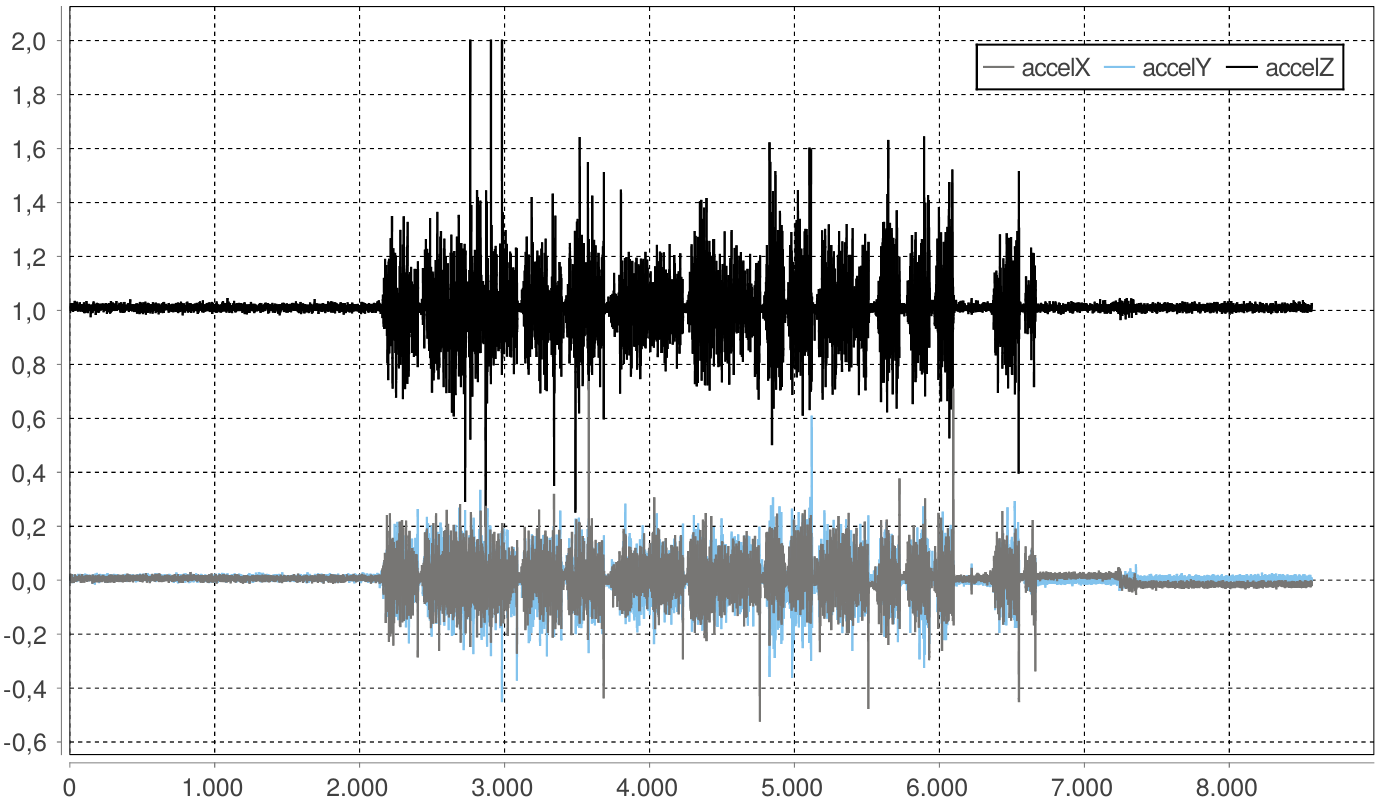}
  \caption{Measurement data excerpt of the acceleration in G $\cdot$ s$^{-1}$}
  \label{fig:measurement-data}
\end{figure}

Next, we used our \emph{modeling environment} (see Fig. \ref{fig:wsn-editor}) to generate executable code for the gateway to which the Nordic Thingy:52\texttrademark{} should connect for forming a \ac{WSN}.
In this case, we were able to generate 100$\,$\% of the gateway code. In addition to that, we generated the static ontology instances according to the \ac{SSN} ontology which were ought to describe the network structure and the contained sensors, see List. \ref{lst:sensor-platform-definition}.
We then operated the robot platform on both smooth and bumpy floor sections.
The measurement data exhibited spikes in the Z-axis of the acceleration graph when the robot operated on a bumpy floor section, see Fig. \ref{fig:measurement-data}.
To monitor the measurement we also implemented a dashboard which is capable of visualizing all captured measurement data.
In addition to that all sensor measurement data was written to \ac{CSV} files and published to a \ac{MQTT} broker.

\begin{figure}
\begin{lstlisting}[caption={Definition excerpt of the sensor platform in OWL/XML}, label={lst:sensor-platform-definition}, frame=single, xleftmargin=3.4pt,xrightmargin=3.4pt, language=XML, morekeywords={rdf:Description, rdf:about, sosa:hosts, rdfs:label, rdf:type},basicstyle=\ttfamily\tiny]
<rdf:Description rdf:about="Thingy52-...">
    <sosa:hosts>NineAxisMotionSensor-...</sosa:hosts>
    <sosa:hosts>LowPowerAccelerometer-...</sosa:hosts>
    <sosa:hosts>BatterySensor-...</sosa:hosts>
    <sosa:hosts>GasSensor-...</sosa:hosts>
    <sosa:hosts>ColorSensor-...</sosa:hosts>
    <sosa:hosts>PressureSensor-...</sosa:hosts>
    <sosa:hosts>HumidityAndTemperatureSensor-...</sosa:hosts>
    <rdfs:label>Nordic Semiconductor Thingy:52 compact multi-sensor prototyping platform</rdfs:label>
    <rdf:type>sosa:Platform</rdf:type>
</rdf:Description>
\end{lstlisting}
\end{figure}
\begin{figure*}
  \centering
  \includegraphics[width=\linewidth]{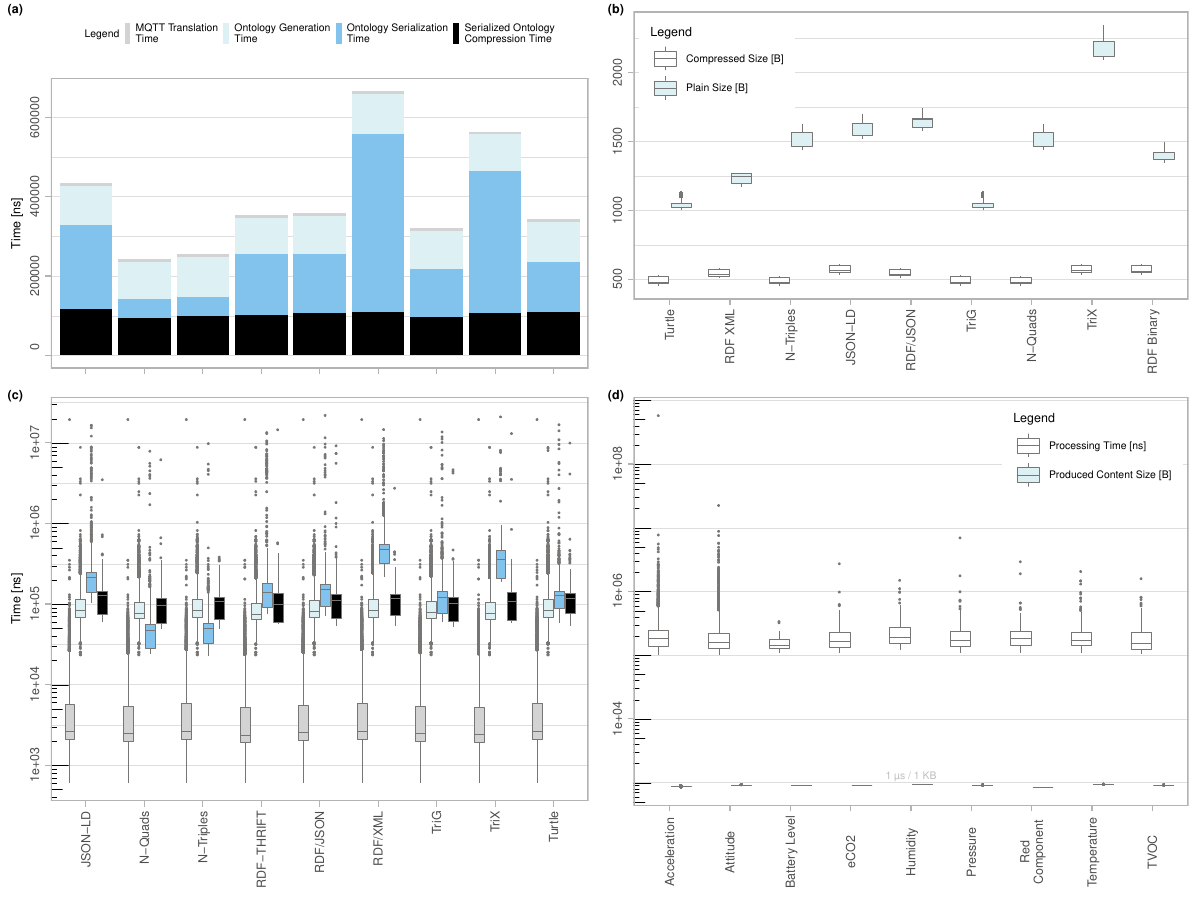} 
  \caption{(a) and (c) depict the execution times of individual micro-ontology generation phases for different \ac{OWL} serialization formats using mean and exact values, respectively. (b) shows the effects of GZIP compression on micro-ontology size for the serialization formats. (d) depicts the processing time and micro-ontology size for each generated property type using Turtle as \ac{OWL} serialization format}
  \label{fig:evaluation-results}
\end{figure*}

While the visualizations and \ac{CSV} files are appropriate when actually performing the measurements, they are not helpful when interpreting measurement data at a later point in time or when sharing the data.

\setlength{\belowcaptionskip}{-10pt}

\begin{figure}
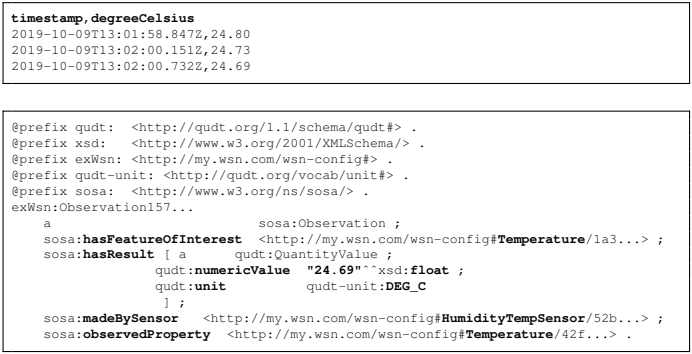

  \subfloat{
\begin{lstlisting}[frame=single,lastline=4,morekeywords={timestamp,degreeCelsius},basicstyle=\ttfamily\tiny]
timestamp,degreeCelsius
2019-10-09T13:01:58.847Z,24.80
2019-10-09T13:02:00.151Z,24.73
2019-10-09T13:02:00.732Z,24.69
\end{lstlisting}}\\
  \subfloat{
\begin{lstlisting}[frame=single,language=Turtle,basicstyle=\ttfamily\tiny]
@prefix qudt:  <http://qudt.org/1.1/schema/qudt#> .
@prefix xsd:   <http://www.w3.org/2001/XMLSchema/> .
@prefix exWsn: <http://my.wsn.com/wsn-config#> .
@prefix qudt-unit: <http://qudt.org/vocab/unit#> .
@prefix sosa:  <http://www.w3.org/ns/sosa/> .
exWsn:Observation157...
    a                          sosa:Observation ;
    sosa:hasFeatureOfInterest  <http://my.wsn.com/wsn-config#Temperature/1a3...> ;
    sosa:hasResult [ a      qudt:QuantityValue ;
                  qudt:numericValue  "24.69"^^xsd:float ;
                  qudt:unit          qudt-unit:DEG_C
                   ] ;
    sosa:madeBySensor   <http://my.wsn.com/wsn-config#HumidityTempSensor/52b...> ;
    sosa:observedProperty  <http://my.wsn.com/wsn-config#Temperature/42f...> .
\end{lstlisting}}
  \caption{CSV (top) versus enriched OWL measurement data (bottom)}
  \label{fig:csv-vs-owl}
\end{figure}

This is where the RDFier component comes in:  In addition to the generation of the gateway code and the static instances from the \ac{SSN} ontology, we used our \ac{EMF} modeling environment to generate the RDFier which is capable of producing and publishing \ac{SSN} ontology instances representing sensor \emph{observations} as \ac{RDF} data.
For a comparison between raw \ac{CSV} data and excerpt of corresponding enriched measured data, see Fig. \ref{fig:csv-vs-owl}.



\section{Evaluation}
\label{sec:evaluation}
In order to evaluate our approach with respect to its overall quality, we validated the correct generation of \frameworkname{} code, as well as the proper serialization of semantically enriched measurement data.
As apparent from Fig. \ref{fig:csv-vs-owl}, our approach enriches raw measurement data with explicit, interoperable semantics due to the utilization of standardized ontologies.
Likewise, our integrated modeling environment is capable of generating not only executable gateway code, but also the static ontology and the RDFier component which are both responsible for the semantic enrichment.

Furthermore, the \frameworkname{} model in our use case could be created in the modeling environment by a person without expert knowledge due to the suitable abstractions provided by the metamodel.
These models can be stored and loaded, supporting \ac{WSN} configuration and management.

As \acp{WSN} are often employed in real-time applications \cite{ali2017comprehensive}, latency is a key performance factor we considered during our evaluation.
To evaluate the impact of our framework on \ac{WSN} latency, we conducted three different measurements on an Intel\textregistered{} Core\texttrademark{} i5-750 CPU at 2.67$\,$GHz with 8$\,$GB RAM.
Data for the first, second, and third measurement were collected by letting the executable run with the same \ac{WSN} configuration for $5$, $60$, and $4$ minutes, collecting over $179\,000$, $266\,000$, and $17\,000$ data points, respectively.
For each measurement, the executable was simply run until an adequate number of data points was acquired.

The goal of the first measurement was to get an insight into the execution times of the individual phases of the micro-ontology generation process, which comprise \ac{MQTT} translation, ontology generation, serialization, and compression, in order.
In particular, we were interested in the performance implications of using different \ac{OWL} serialization formats and the effects of GZIP compression on generation time.

Fig. \ref{fig:evaluation-results} (a) and (c) show the results of the first measurement, which imply a non-significant difference in generation times between the individual \ac{OWL} serialization formats.
The majority of generation is performed within 500$\,$\textmu s, with some outliers exceeding 1$\,$ms, which can be most likely explained by garbage collection side effects of the Java Virtual Machine.

Consequently, we can draw the conclusion that any overhead introduced by our approach in terms of latency is going to be dominated by the transmission of the ontology instances over the network, and not by the generation of the dynamic ontology instances.
This is a direct result of the considerations described in \mbox{Section \ref{sec:system-architecture}}, in which we essentially had to make a trade-off between non-intrusiveness and latency, favoring the former over the latter.
It is obvious that optimal latency would be achieved by integrating the RDFier component directly on the server hosting the network broker as depicted in Fig. \ref{fig:system-architecture}, but in this case full access to the server would be required which is not always possible.

With the second measurement, we intended to measure the effects of GZIP compression on micro-ontology size for the serialization formats used in the first measurement.
The results of this measurement are shown in Fig. \ref{fig:evaluation-results} (b).
While the achievable compression ratio is highly dependent on the chosen serialization format, the inter-format variance between the compressed micro-ontology sizes is negligible.
This low variance in compressed size allows for a high degree of freedom with respect to serialization format selection. 

With the third measurement, we intended to analyze the micro-ontology generation time in relation to the generated micro-ontology size.
Fig. \ref{fig:evaluation-results} (d) shows the results of the measurement for each generated property type using Turtle as \ac{OWL} serialization format.
The micro-ontology instances generated by the RDFier all have sizes less than 1$\,$KB and non-significant inter-property size variance, due to our provisioning of semantic information in static and dynamic form, as shown in Fig. \ref{fig:approach-overview} \circled{\scriptsize{S2}} and \circled{\scriptsize{R4}}, respectively.

\section{Conclusion and Future Work}
\label{sec:conclusion}
We have presented a model-based approach to the configuration and generation of executable code for \acp{WSN} which includes a framework for an ontology-based semantic enrichment and provisioning of measurement data.

The current version of the \ac{WSN} metamodel does not capture the entirety of the SSN ontology.
On the one hand, this is due to limitations in the ontology itself.
For instance the currently published set of \acl{QUDT} (\acs{QUDT}) ontologies does not yet contain definitions for the representation of specialized vectors such as those representing the attitude of a sensor platform.

On the other hand, the mapping of the SSN ontology to \ac{EMF} is challenging due to differences in the expressiveness of the respective (meta-)metamodels.
For example, multiple classification for class instances is difficult to represent in Ecore.
Bridging this gap is non-trivial and will be a major part of future efforts.
One possible solution could be the provision of generic mechanisms that allow the definition of such ontological constructs within the modeling environment.
Such an approach could potentially be generalizable to arbitrary ontologies.

While the current tree-based implementation of the modeling environment is functional, a graphical modeling environment using suitable visualizations can further improve usability.
The code generation framework could be extended to allow the generation of server configuration code.
For commercial off-the-shelf gateways, the sensor data interpretation \ac{DSL} can be extended to support industry-standard data exchange formats such as JSON or XML.
Another desirable feature would be the possibility of live network management using the corresponding \frameworkname{} model.

Our approach paves the way for significant reductions of \ac{WSN} setup effort and management complexity by generating 100$\,$\% of \frameworkname' application code. In addition to that, our approach accelerates the transition to general device interoperability in IoT with regard to measurement data provisioning.
Furthermore, our findings indicate a significant reduction in \ac{MQTT} payload sizes by compressing the micro-ontology instances.
Future work will evaluate the incorporation of advanced delta-based techniques for network traffic reduction into our approach.

\section*{Acknowledgment}
This paper is partially funded by the BMBF within the project COMPACT (grant number 01\textbar{}S17028C), and by the BMWi within the project InsightProducts (grant number 228EN/2).


%
%

\balance

\bibliographystyle{IEEEtran}
\bibliography{main}

\end{document}